\documentclass[a4paper,11pt]{article}
\usepackage{pos}

\usepackage{amssymb}
\usepackage{braket}
\usepackage{siunitx}
\usepackage{booktabs}

\graphicspath{{Fig/}{../QC_BESIII_ICHEP2026/Fig/}}

\newlength{\panelheight}
\newcommand{\ifb}{\ensuremath{~\mathrm{fb}^{-1}}}

\newcommand{\DtoKSpipi}{D\!\to K_S\pi^+\pi^-}
\newcommand{\DtoKKpipi}{D\!\to K^+K^-\pi^+\pi^-}

\newcommand{\BtoDK}{B^{\pm}\!\to DK^{\pm}}

\newcommand{\ee}{e^+e^-}
\newcommand{\DD}{D\bar{D}}
\newcommand{\KSL}{K_{S,L}}

\title{Quantum-correlation of neutral charmed mesons at BESIII}
\ShortTitle{Quantum-correlation of neutral charmed mesons at BESIII}

\author*[a]{Shenghui Zeng}
\onbehalf{on behalf of the BESIII Collaboration}

\affiliation[a]{H.~H.~Wills Physics Laboratory, University of Bristol,\\
  Tyndall Avenue, Bristol BS8 1TL, United Kingdom}

\emailAdd{shenghui.zeng@cern.ch}

\abstract{%
Quantum-correlated $D\bar D$ pairs give direct access to the strong-phase differences between $D^0$ and $\bar D^0$ decays, which are essential inputs to the CKM angle $\gamma$ and charm mixing. The BESIII experiment has accumulated the largest data sample at the $\psi(3770)$ resonance, corresponding to an integrated luminosity of $20.3\,\text{fb}^{-1}$. The neutral $D\bar D$ pairs produced at $\psi(3770)$ are in a $C$-odd state. We report new and improved measurements of the strong-phase differences in $D\rightarrow K_{S,L}\pi^+\pi^-$, $D\rightarrow K^+K^-\pi^+\pi^-$, $D\rightarrow K^-\pi^+\pi^+\pi^-$ and $D\rightarrow K^-\pi^+\pi^0$ decays, alongside a measurement of the $CP$-even fraction in $D\rightarrow K^+K^-\pi^+\pi^-$. Furthermore, we present a novel method that studies the $D\bar D$ quantum correlations using data collected at centre-of-mass energies from $4.13$ to $4.23\,$GeV, where the $D\bar D$ pairs are produced in $C$-even or $C$-odd states, according to the number of accompanying photons. This gives the first observation of $C$-even $D\bar D$ pairs and a new measurement of the strong-phase difference in $D\rightarrow K^-\pi^+$ decays.}

\FullConference{43rd International Conference on High Energy Physics (ICHEP 2026)\\
30 July -- 5 August, 2026\\
Natal, Brazil\\}

\begin{document}
\maketitle

%
%
%
%

\section{Introduction}
\label{sec:intro}


The Cabibbo-Kobayashi-Maskawa (CKM) angle $\gamma\equiv\arg(-V_{ud}V_{ub}^*/V_{cd}V_{cb}^*)$ is one of the least constrained angles and is the only angle of the Unitarity Triangle that can be measured in tree-level decays.
Its direct determination, $(62.8\pm2.6)^\circ$~\cite{LHCb:2025comb}, is therefore a benchmark for the indirect determinations that involve loop processes, and the comparison between them can probe contributions from new physics.
Such a direct measurement uses tree-level \mbox{$B^\pm\to DK^\pm$} decays, in which $D$ is a superposition of $D^0$ and $\bar D^0$ that decays to a common final state $f$.
The interference depends on $\gamma$ and the strong-phase difference between
the $D^0\to f$ and $\bar D^0\to f$ amplitudes. The strong-phase differences also enter
measurements of charm mixing and $CP$ violation, such as the observation of a
non-zero mass difference between the neutral charm-meson
eigenstates~\cite{LHCb:2021mixing}.

Quantum-correlated $\DD$ pairs give access to the strong-phase differences directly.
Hereafter, $D$ without superscript is a neutral $D$ of unspecified flavour, and charge conjugation is implied.
At the $\psi(3770)$, the neutral $D\overline{D}$ pair is produced in a $C$-odd state, so the determination of the quantum state of one $D$ meson fixes the state of the other.
The BESIII experiment has recorded the world's largest data sample, $20.3\ifb$, at the $\psi(3770)$ resonance.
Above the $D^*\bar{D}$ threshold, BESIII has collected a data sample of $7.13\ifb$ at centre-of-mass energies of $4.13$ to $4.23\,$GeV, where the $\DD$ pairs come from $D^*\bar{D}^{(*)}$ production and so are in both $C$-even and $C$-odd states.
These data samples make BESIII a unique laboratory to study the strong-phase differences.

%

\section{Strong-phase differences in $D\rightarrow K_{S,L}\pi^+\pi^-$ decays}
\subsection{Updated model-independent binned measurement}
\label{sec:kspipi}
The strong phase of $\DtoKSpipi$ varies significantly across the Dalitz plot, which makes this decay the most sensitive $D$ mode for $\gamma$ and the charm-mixing parameter measurements.
The BESIII experiment has measured the strong-phase parameters $(c_i, s_i)$ with a data sample of $7.93\ifb$ in three binning schemes with $2\times8$ bins~\cite{BESIII:2025kspipi}.
The $D\to K_L\pi^+\pi^-$ decay serves as both a signal and a tag mode in this study.
The earlier study linked the $(c'_i, s'_i)$ in $D\to K_L\pi^+\pi^-$ with $(c_i, s_i)$ by introducing a difference from amplitude model prediction~\cite{BESIII:2020kspipi}, which brings a weak model assumption here.
In this analysis, a larger data sample is used and both results with and without this model assumption are presented, both provide the most precise measurement to date. In Fig.~\ref{fig:kspipi} (a) and (b), the optimal binning scheme and the result are shown.
With the study of $\BtoDK$, $\DtoKSpipi$ simulation samples that have negligible $B$ statistical uncertainty,
the propagated uncertainty from the $(c_i, s_i)$ measured from optimal binning scheme to $\gamma$ is found to be $0.9^{\circ}$ ($1.5^{\circ}$) based on constrained (unconstrained) results.
The first $\gamma$ measurement with the upgraded LHCb detector~\cite{LHCb:2026gamma} has already used the parameters without the model assumption, so its $\DtoKSpipi$ input is free from amplitude-model dependence.
\begin{figure}[t]
  \centering
  \includegraphics[height=\panelheight]{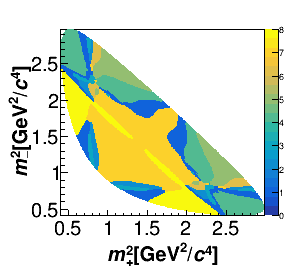}\hfill
  \includegraphics[height=\panelheight]{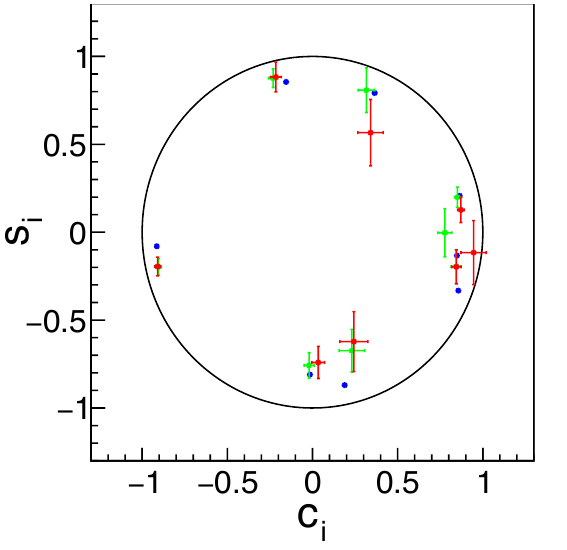}\hfill
  \includegraphics[height=\panelheight]{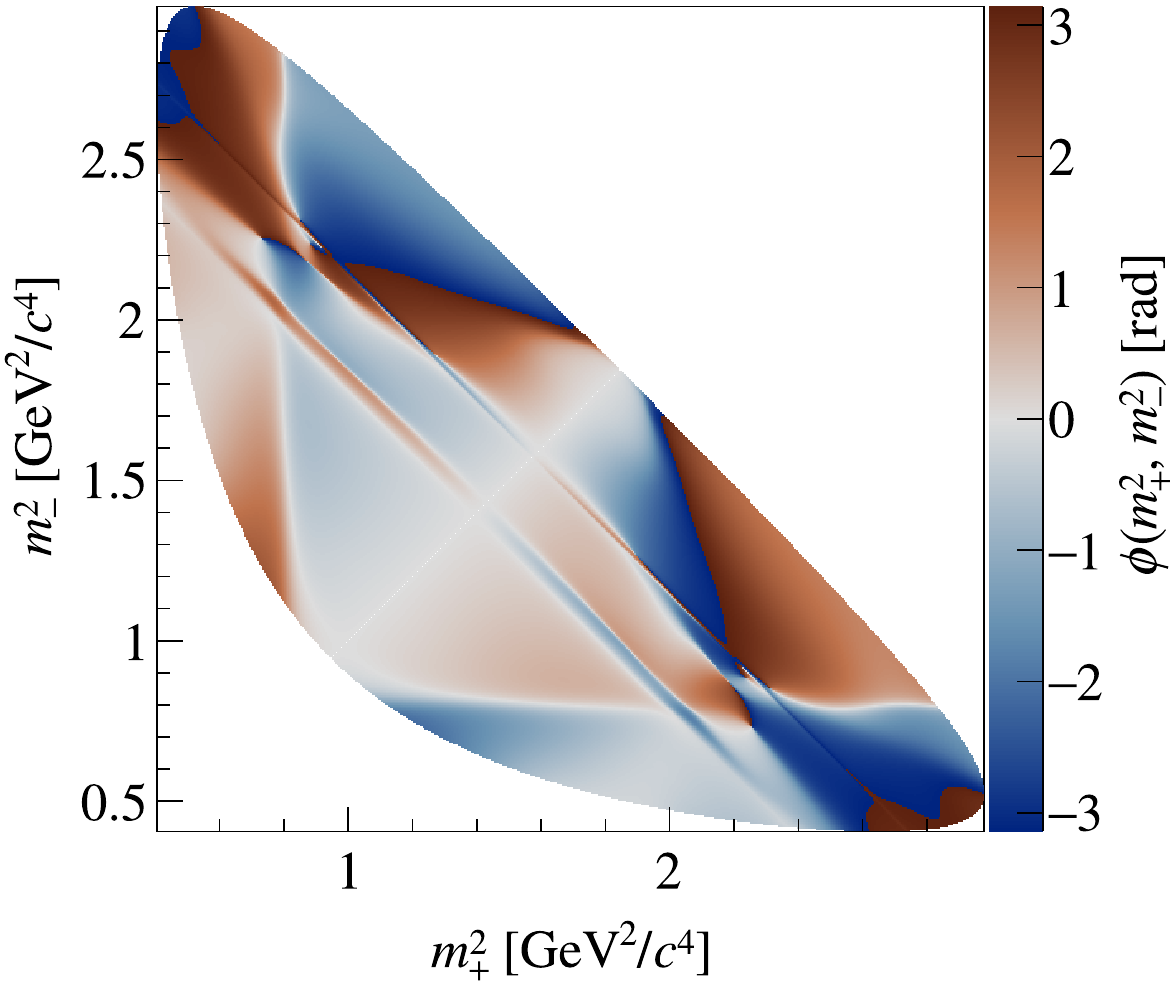}\hfill
  \includegraphics[height=\panelheight]{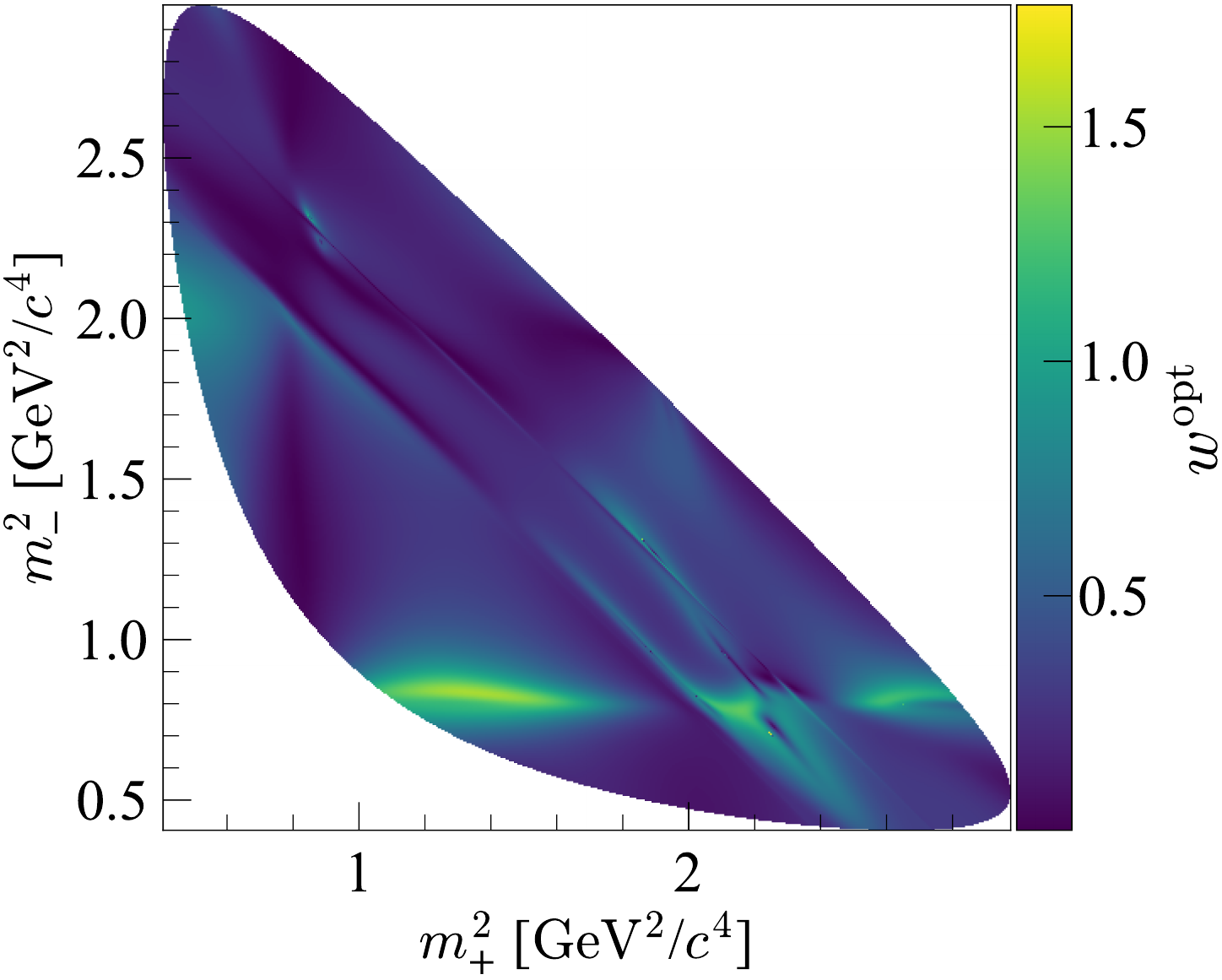}
  \caption{For $\DtoKSpipi$ strong-phase analysis, from left to right: (a)~Optimal
    binning scheme ($m^2_\pm\equiv m^2(K_S\pi^\pm)$), and (b)~$(c_i,s_i)$ in this
    scheme from $7.93\ifb$, without (red) and with (green) the model link to $K_L\pi^+\pi^-$, and the model prediction
    (blue)~\cite{BESIII:2025kspipi}. (c)~Strong-phase difference
    $\phi(z)$ and (d)~optimal weight $w^{\rm opt}(z)$, which define the
    per-event weights~\cite{BESIII:2026unbinned1}.}
  \label{fig:kspipi}
\end{figure}

\subsection{Unbinned measurement: determination of optimal Fourier weights}
\label{sec:unbinned}

Binning averages the interference inside each bin, and the current binning schemes keep about 85\% of the statistical sensitivity to $\gamma$.
The joint BESIII and LHCb analysis uses the \textit{optimal Fourier} weights $w^{\text{opt}}\cos[k\phi(z)]$ and $w^{\text{opt}}\sin[k\phi(z)]$~\cite{Poluektov2018}.
Here, $\phi(z)$ is the strong-phase difference at a Dalitz plot point $z$, and the optimal weights $w^{\text{opt}}$ include the amplitude ratio, efficiency and background effects. The Dalitz plot distribution of $\phi(z)$ and $w^{\text{opt}}$ are shown in Fig.~\ref{fig:kspipi} (c) and (d) respectively.
Both optimal weight and $\phi(z)$ are predicted from an amplitude model, however, the weighted integrals are determined from data.
A poor model only dilutes the sensitivity but does not bias the result itself.
At BESIII, these weighted integrals have been measured with $8\ifb$ $D\rightarrow K_{S,L}h^+h^-$ decays.
LHCb can then use this information to measure the weighted \mbox{$B^{\pm}\rightarrow Dh^{\pm}$} $(h=K,\pi)$ yields with $9\ifb$ of $pp$ collision data.
A simultaneous fit with Fourier expansion $k\leq2$ for $D\rightarrow K_S\pi^+\pi^-$ and $k\leq1$ for $D\rightarrow K_SK^+K^-$ gives $\gamma=(71.3\pm5.0)^{\circ}$~\cite{BESIII:2026unbinned1} (Fig.~\ref{fig:gamma_inputs}(a)), which is the most precise single measurement to date.
Compared with the previous measurement that uses the same LHCb data~\cite{LHCb:2021kshh}, this measurement improves the statistical precision on $\gamma$ by 5\%, and reduces the uncertainty due to the BESIII input to $0.6^{\circ}$.


\section{First model-independent strong-phase measurement of $D\to K^+K^-\pi^+\pi^-$ decay}
\label{sec:kkpipi}
The $D\to K^+K^-\pi^+\pi^-$ decay has been studied by the LHCb Collaboration and found to provide interesting sensitivity to $\gamma$~\cite{LHCb:2023kkpipi}.
Similar to the $D\rightarrow K_S\pi^+\pi^-$ decay, the strong-phase differences of $D\to K^+K^-\pi^+\pi^-$ decay can be measured in phase-space bins.
The binning has been done by defining: $\Delta\delta_D \equiv \arg\left(\mathcal{A}_{D^0}/\mathcal{A}_{\bar{D}^0}\right), r_D \equiv \left|\mathcal{A}_{D^0}/\mathcal{A}_{\bar{D}^0}\right|$,
and therefore the five-dimensional phase space can be projected into a two-dimensional surface. 
With $20.3\ifb$ of $\psi(3770)\rightarrow D\bar D$ data, the first model-independent measurement of the binned strong-phase difference $(c_i, s_i)$ for $D\to K^+K^-\pi^+\pi^-$ has been reported~\cite{BESIII:2025kkpipi} (Fig.~\ref{fig:gamma_inputs}(b)), this is consistent with the model predicted value.
This analysis also gives the $CP$-even fraction $F_+=0.754\pm0.010\pm0.008$, where the first uncertainty is statistical and the second is systematic.
For a $B$ sample with a size equivalent to LHCb Run 1 and Run 2, the contribution to $\gamma$ is about $10^{\circ}$, as a reference the statistical uncertainty of the $\gamma$ itself is $12.7^{\circ}$.
The LHCb experiment reported its first model-independent
binned analysis of $B^\pm\to[K^+K^-\pi^+\pi^-]_Dh^\pm$ and
$B^\pm\to[\pi^+\pi^-\pi^+\pi^-]_Dh^\pm$ decays using this result, and gives
$\gamma=(53.9^{+9.5}_{-8.9})^\circ$~\cite{LHCb:2026kkpipi}.

\section{Coherence factors and strong-phase differences in $D\rightarrow K^{\pm}n\pi\ (n\geq2)$ decays}

\label{sec:knpi}
The $D\to K^-\pi^+\pi^+\pi^-$ decay and $D\to K^-\pi^+\pi^0$ decay are an important category of charm decays and they provide unique sensitivity for $\gamma$ measurement.
These quasi-flavour decay modes combine both Cabibbo-favoured and doubly Cabibbo-suppressed amplitudes,
and the interference can be measured as $R_fe^{i\delta^f_D}$. Here, $R_f$ is the coherence factor and $\delta_D^f$ is the strong-phase difference for decay mode $f$.
With $7.93\ifb$ of quantum-correlated $D\bar{D}$ data, BESIII obtains $R_{K3\pi}=0.51\pm0.04$,
$\delta_D^{K3\pi}=(182^{+14}_{-13})^\circ$, $R_{K\pi\pi^0}=0.75\pm0.03$ and
$\delta_D^{K\pi\pi^0}=(209^{+7}_{-8})^\circ$~\cite{BESIII:2026knpi}. The
uncertainty on $R_{K3\pi}$ is two to three times smaller than in the CLEO-c
and earlier BESIII results (Fig.~\ref{fig:gamma_inputs}(c,d)). The
$K^-\pi^+\pi^+\pi^-$ parameters are also measured in four phase-space bins,
this is benefiting from the large local coherence effect up to $0.80^{+0.07}_{-0.06}$, from the overlapping resonance contributions.
This enables a stand-alone
determination of $\gamma$. With the $9\ifb$ LHCb sample, its expected
uncertainty is $8.5^\circ$, of which $3.5^\circ$ comes from the BESIII inputs.


\begin{figure}[htb]
  \centering
  \includegraphics[height=\panelheight]{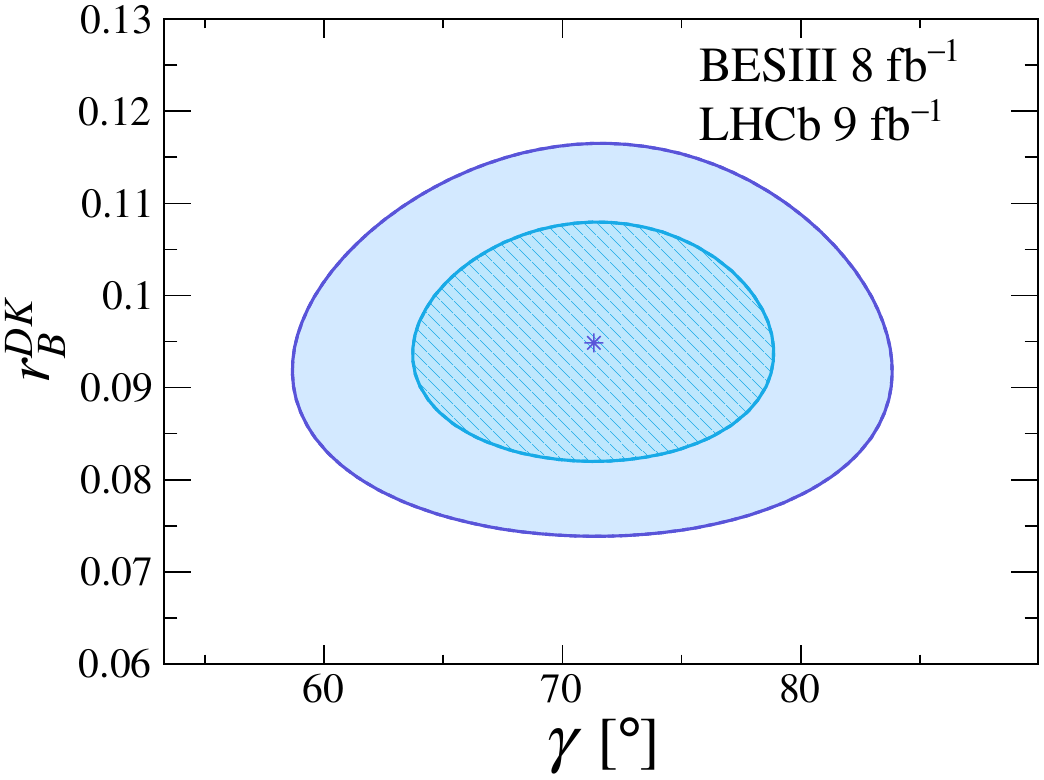}\hfill
  \includegraphics[height=\panelheight]{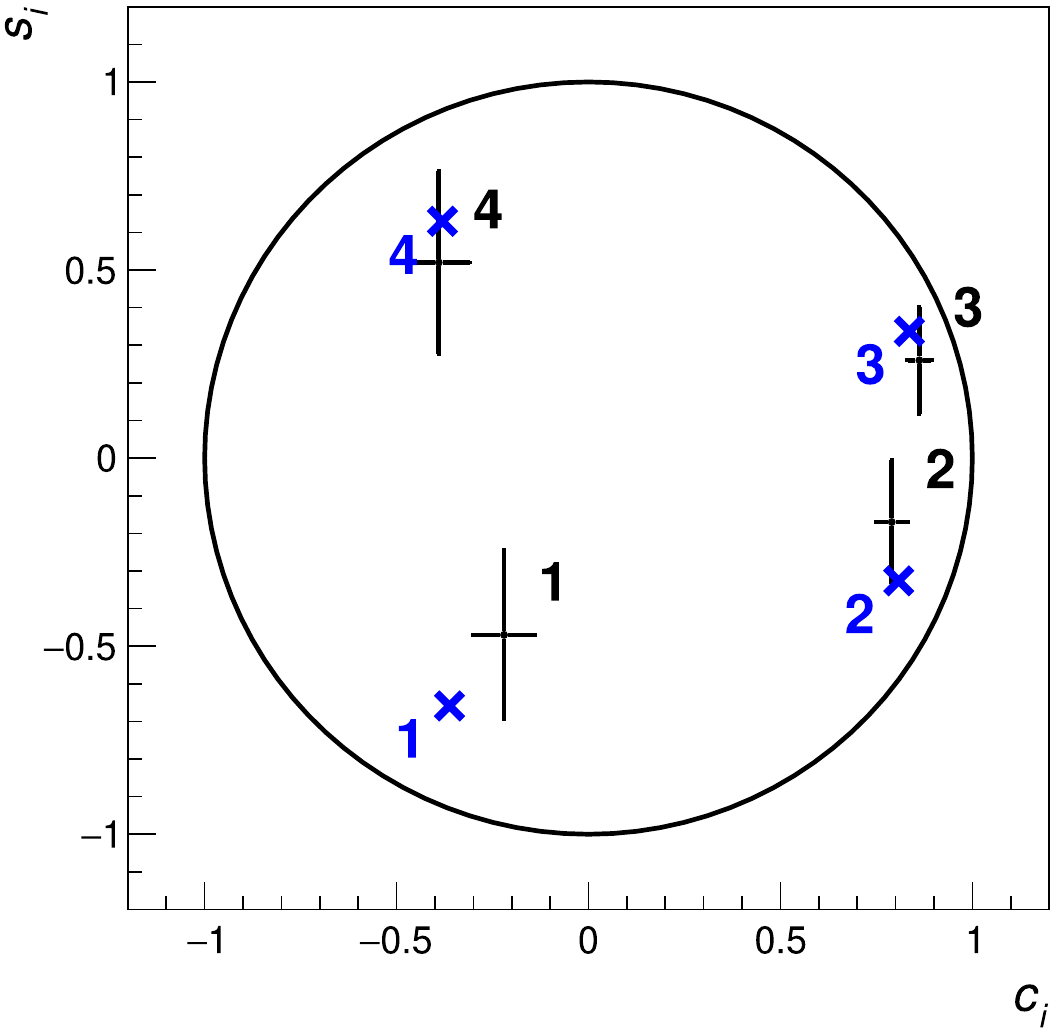}\hfill
  \includegraphics[height=\panelheight]{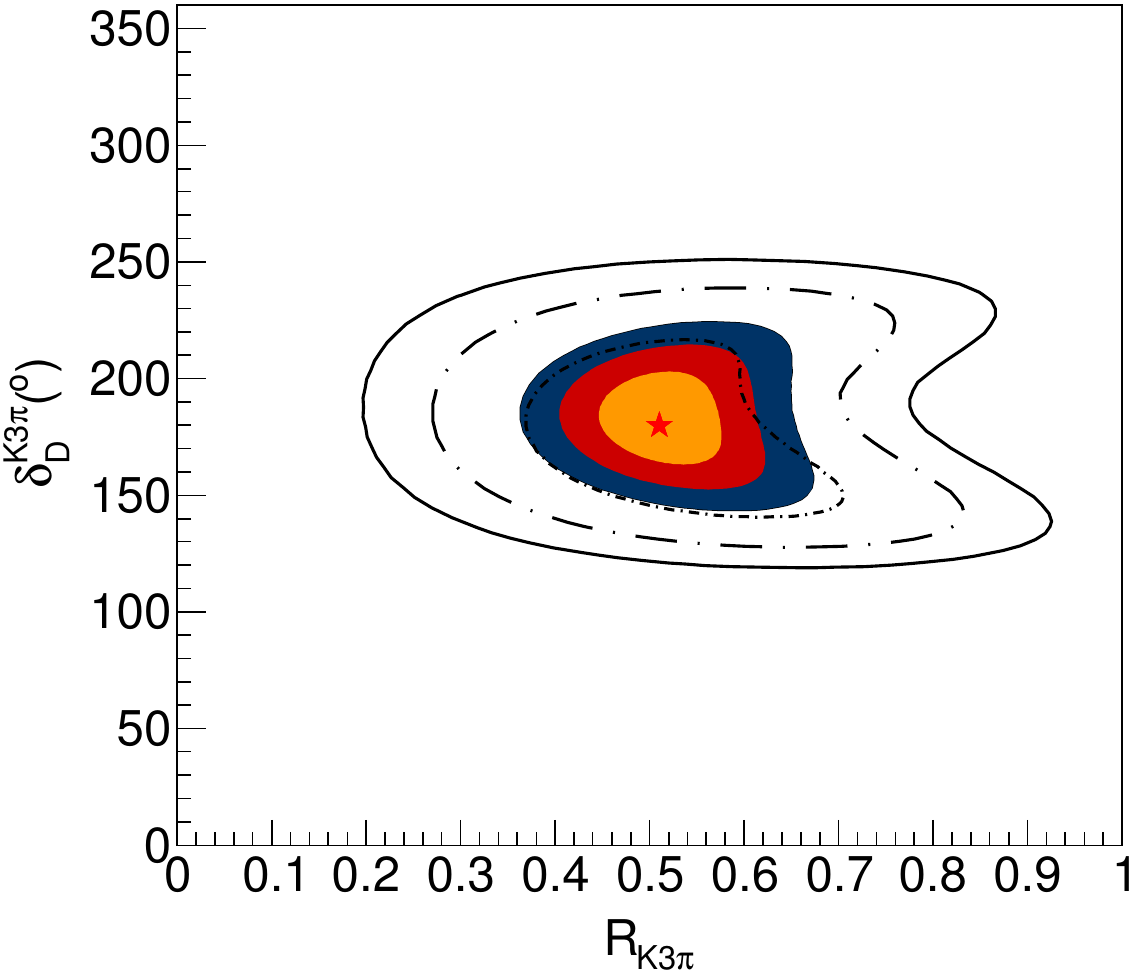}\hfill
  \includegraphics[height=\panelheight]{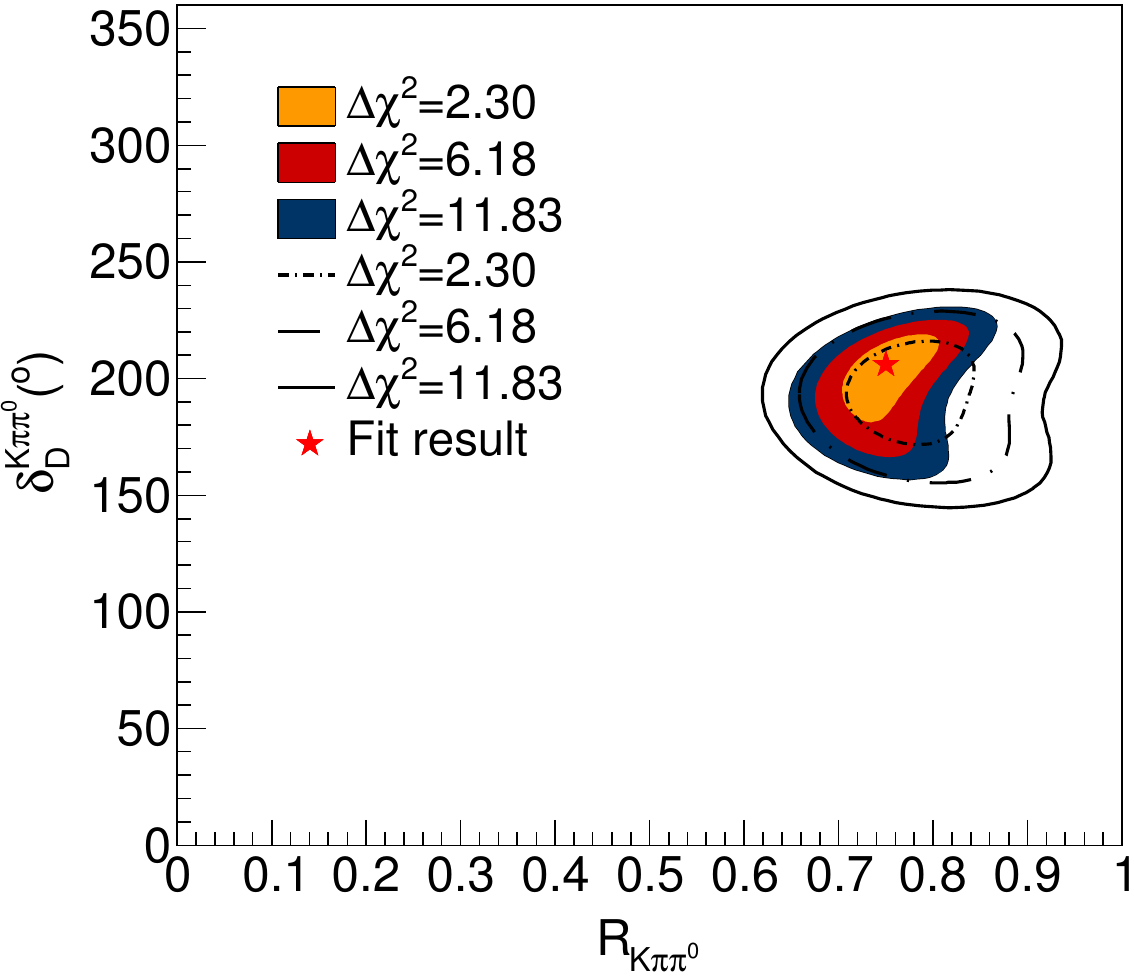}
  \caption{From left to right: (a)~two-dimensional 68\% and 95\% confidence-level (CL) contours of
    $(\gamma,r_B^{DK})$ from the joint BESIII and LHCb unbinned
    analysis~\cite{BESIII:2026unbinned1}. (b)~Measured $(c_i,s_i)$ of
    $\DtoKKpipi$ in the four bin pairs (black, 68\% CL, statistical only)
    compared with the amplitude-model prediction (blue
    crosses)~\cite{BESIII:2025kkpipi}. $(R_f,\delta^f_D)$, for decay modes $f$ of
    (c)~$D\to K^-\pi^+\pi^+\pi^-$ and (d)~$D\to K^-\pi^+\pi^0$, from the
    global fit.
    The black lines show the earlier $2.93\ifb$
    result~\cite{BESIII:2026knpi}.}
  \label{fig:gamma_inputs}
\end{figure}

\section{$C$-even $D\bar D$ above threshold: a new source of correlated pairs}
\label{sec:ceven}
On and above the $D^*\bar D$ threshold, $\DD$ pairs are produced with photons and neutral pions.
When these consist of $n$ photons or any number of the $\pi^0$, the $\DD$ pairs are then expected to have an eigenvalue of $C=(-1)^{n+1}$~\cite{BESIII:2025cevenPRL}.
The BESIII experiment fully reconstructed both $D$ mesons and explored five different production mechanisms, using $7.13\ifb$ data sample at $\sqrt s=4.13$--$4.23\,$GeV, for the quantum-correlation studies.
The productions of $\DD$, $\pi^0\DD$, $\gamma\gamma/\pi^0\pi^0\DD$ with $C=-1$, and $\gamma\DD$, $\gamma\pi^0\DD$ with $C=+1$ are studied with the missing energy and mass. 
As shown in Fig.~\ref{fig:ceven_qc}, the efficiency-corrected yields match the prediction well, which leads to the first observation of the $C$-even quantum-correlation in the $\ee\to X\DD$ processes~\cite{BESIII:2025cevenPRL,BESIII:2025cevenPRD}.

\begin{figure}[htb]
  \centering
  \includegraphics[width=0.75\textwidth]{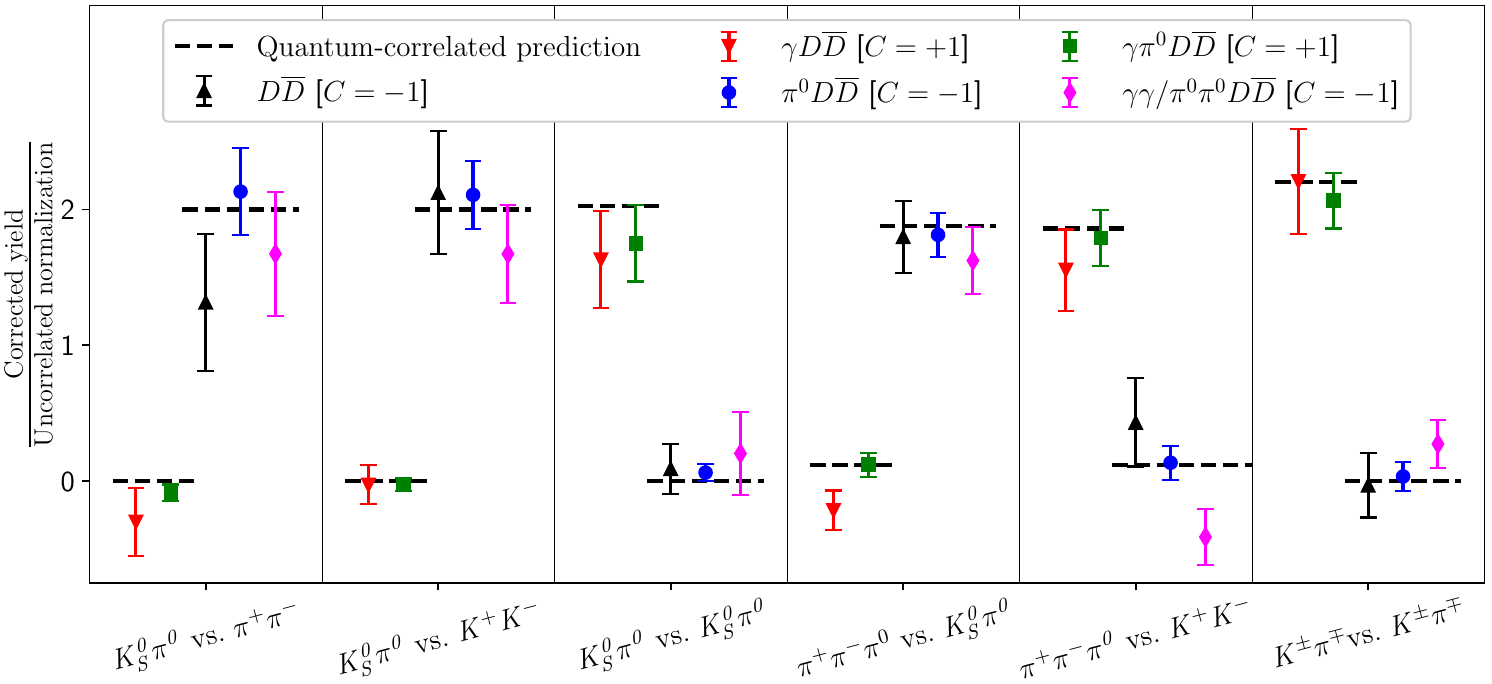}
  \caption{The ratios of efficiency-corrected yields observed in data to those
    expected in the absence of correlations and mixing for each $\DD$ final
    state originating from each production mechanism. The displayed errors
    include systematic uncertainties~\cite{BESIII:2025cevenPRL}.}
  \label{fig:ceven_qc}
\end{figure}

The strong-phase differences $\delta_D^{K\pi}$ can then be measured with these data set, giving $\delta_D^{K\pi} = (192.8^{+11.0+1.9}_{-12.4-2.4})^{\circ}$, where the first uncertainty is statistical and the second is systematic~\cite{BESIII:2025cevenPRD}.
This result agrees with the previous BESIII measurement~\cite{BESIII:2022kpi}, with a three times smaller systematic uncertainty, and the combination of the two measurements is $\delta_D^{K\pi} = (189.2^{+6.9+3.4}_{-7.4-3.8})^{\circ}$.
Moreover, the $C$-even decay rates have a linear relation to the mixing parameters $x$ and $y$, which gives a promising sensitivity for the time-integrated mixing measurements at $e^+e^-$ colliders.
The study with a predicted $1\,{\rm ab}^{-1}$ data at each of four energies from $4.01$ to $4.26\,$GeV at a future super tau-charm facility gives $\sigma(x)\approx5.6\times10^{-4}$ and $\sigma(y)\approx1.5\times10^{-4}$~\cite{BESIII:2025cevenPRL}, comparable to the present world averages.


\section{Summary}
\label{sec:summary}

\begin{table}[t]
  \centering
  \caption{Results reviewed here. $\sigma_\gamma$ is the contribution of the
    BESIII inputs to the uncertainty on $\gamma$, as estimated in each paper.}
  \label{tab:summary}
  \small
  \begin{tabular}{@{}llcc@{}}
    \toprule
    Mode & Observable & Integrated luminosity ($\mathrm{fb}^{-1}$) & $\sigma_\gamma$\\
    \midrule
    $D\to\KSL\pi^+\pi^-$ & $(c_i,s_i)$ in $3\times(2\times8)$ bins & $7.93$ & $0.7^\circ$--$1.5^\circ$\\
    $D\to\KSL h^+h^-$ & Optimal Fourier weights & $8$ (+LHCb $9$) & $0.6^\circ$\\
    $D\to K^+K^-\pi^+\pi^-$ & $(c_i,s_i)$ in $2\times4$ bins, $F_+$ & $20.3$ & ${\approx}10^\circ$\\
    $D\to K^-\pi^+\pi^+\pi^-$ & $R$, $\delta_D$ (global and in 4 bins) & $7.93$ & $3.5^\circ$\\
    $D\to K^-\pi^+\pi^0$ & $R$, $\delta_D$ & $7.93$ & --\\
    $D\to K^-\pi^+$ & $\delta_D$ & $7.13$ & --\\
    \bottomrule
  \end{tabular}
\end{table}

The BESIII experiment has provided important $D$ strong-phase inputs to $CP$-violation
studies, most notably the determination of the CKM angle $\gamma$. Novel methods
such as the unbinned strong-phase parametrisation and the exploitation of $C$-even
correlations in $D^{*}\bar D^{(*)}$ data significantly improve both statistical
and systematic uncertainties.
A summary of the results quoted in this report is shown in Tab.~\ref{tab:summary}.
Future measurements with improved analysis methods and increased datasets are expected to push the precision further.



\begin{thebibliography}{99}

  \bibitem{LHCb:2025comb}
  LHCb Collaboration,
  ``Simultaneous determination of the CKM angle $\gamma$ and parameters
  related to mixing and $CP$ violation in the charm sector``,
  LHCb-CONF-2025-003,
  \href{https://doi.org/10.17181/CERN.GFHE.LM6W}{doi:10.17181/CERN.GFHE.LM6W}.

  \bibitem{LHCb:2021mixing}
  LHCb Collaboration,
  ``Observation of the mass difference between neutral charm-meson
  eigenstates``,
  \href{https://doi.org/10.1103/PhysRevLett.127.111801}{Phys. Rev. Lett.
  \textbf{127} (2021) 111801}.

  \bibitem{BESIII:2025kspipi}
  BESIII Collaboration,
  ``Updated model-independent measurement of the strong-phase differences
  between $D^0$ and $\bar D^0\to K^0_{S/L}\pi^+\pi^-$ decays,''
  \href{https://doi.org/10.1007/JHEP06(2025)086}{JHEP \textbf{06} (2025) 086}.

  \bibitem{BESIII:2020kspipi}
  BESIII Collaboration,
  ``Model-independent determination of the relative strong-phase difference
  between $D^0$ and $\bar D^0\to K^0_{S,L}\pi^+\pi^-$ and its impact on the
  measurement of the CKM angle $\gamma/\phi_3$,''
  \href{https://doi.org/10.1103/PhysRevD.101.112002}{Phys. Rev. D \textbf{101} (2020) 112002};
  BESIII Collaboration,
  ``Determination of strong-phase parameters in $D\to K^0_{S,L}\pi^+\pi^-$,''
  \href{https://doi.org/10.1103/PhysRevLett.124.241802}{Phys. Rev. Lett. \textbf{124} (2020) 241802}.

  \bibitem{LHCb:2026gamma}
  LHCb Collaboration,
  ``Measurement of $\gamma$ using $B^{\pm}\to DK^{\pm}$ and $B^{\pm}\to D\pi^{\pm}$
  decays with $D\to K^0_S\pi^+\pi^-$ and $D\to K^0_S K^+K^-$,''
  \href{https://doi.org/10.1007/JHEP08(2026)037}{JHEP \textbf{08} (2026) 037}.

  \bibitem{BESIII:2026unbinned1}
  BESIII and LHCb Collaborations,
  ``Measurement of the CKM angle $\gamma$ in
  \mbox{$B^{\pm}\to D(\to K^0_S h'^+h'^-)h^{\pm}$} decays with a novel approach,''
  Phys. Rev. D (accepted),
  \href{https://arxiv.org/abs/2604.05701}{arXiv:2604.05701};
  BESIII and LHCb Collaborations,
  ``Precise measurement of the CKM angle $\gamma$ with a novel approach,''
  Phys. Rev. Lett. (accepted),
  \href{https://arxiv.org/abs/2604.05712}{arXiv:2604.05712}.

  \bibitem{Poluektov2018}
  A.~Poluektov,
  ``Unbinned model-independent measurements with coherent admixtures of
  multibody neutral $D$ meson decays,''
  \href{https://doi.org/10.1140/epjc/s10052-018-5599-1}{Eur. Phys. J. C \textbf{78} (2018) 121}.

  \bibitem{LHCb:2021kshh}
  LHCb Collaboration,
  ``Measurement of the CKM angle $\gamma$ in $B^{\pm}\to DK^{\pm}$ and
  $B^{\pm}\to D\pi^{\pm}$ decays with $D\to K^0_S h^+h^-$,''
  \href{https://doi.org/10.1007/JHEP02(2021)169}{JHEP \textbf{02} (2021) 169}.

  \bibitem{LHCb:2023kkpipi}
  LHCb Collaboration,
  ``A study of $CP$ violation in the decays $B^{\pm}\to[K^+K^-\pi^+\pi^-]_D h^{\pm}$
  ($h=K,\pi$) and $B^{\pm}\to[\pi^+\pi^-\pi^+\pi^-]_D h^{\pm}$,''
  \href{https://doi.org/10.1140/epjc/s10052-023-11560-5}{Eur. Phys. J. C \textbf{83} (2023) 547}.

  \bibitem{BESIII:2025kkpipi}
  BESIII Collaboration,
  ``Measurement of the strong-phase difference between $D^0$ and
  \mbox{$\bar D^0\to K^+K^-\pi^+\pi^-$} in bins of phase space,''
  \href{https://doi.org/10.1103/PhysRevD.112.012015}{Phys. Rev. D \textbf{112} (2025) 012015}.

  \bibitem{LHCb:2026kkpipi}
  LHCb Collaboration,
  \emph{``A model-independent measurement of the CKM angle $\gamma$ in the decays
  $B^\pm\to[K^+K^-\pi^+\pi^-]_Dh^\pm$ and $B^\pm\to[\pi^+\pi^-\pi^+\pi^-]_Dh^\pm$
  ($h=K,\pi$)``},
  \href{https://doi.org/10.1007/JHEP01(2026)062}{JHEP \textbf{01} (2026) 062}.

  \bibitem{BESIII:2026knpi}
  BESIII Collaboration,
  ``Improved measurements of the coherence factors and strong-phase
  differences in $D\to K^-\pi^+\pi^+\pi^-$ and $D\to K^-\pi^+\pi^0$ with
  quantum-correlated $D\bar D$ decays,''
  \href{https://doi.org/10.1007/JHEP08(2026)038}{JHEP \textbf{08} (2026) 038}.

  \bibitem{BESIII:2025cevenPRL}
  BESIII Collaboration,
  ``First observation of quantum correlations in $e^+e^-\to XD\bar D$ and
  $C$-even constrained $D\bar D$ pairs,''
  \href{https://doi.org/10.1103/dkgg-9twj}{Phys. Rev. Lett. \textbf{135} (2025) 171901}.

  \bibitem{BESIII:2025cevenPRD}
  BESIII Collaboration,
  ``Novel measurement of the strong-phase difference between \mbox{$D^0\to K^-\pi^+$}
  and \mbox{$\bar D^0\to K^-\pi^+$} decays using $C$-even and $C$-odd
  quantum-correlated $D\bar D$ pairs,''
  \href{https://doi.org/10.1103/q6qr-8csr}{Phys. Rev. D \textbf{112} (2025) 072006}.

  \bibitem{BESIII:2022kpi}
  BESIII Collaboration,
  ``Improved measurement of the strong-phase difference $\delta_D^{K\pi}$ in
  quantum-correlated $D\bar D$ decays,''
  \href{https://doi.org/10.1140/epjc/s10052-022-10872-2}{Eur. Phys. J. C \textbf{82} (2022) 1009}.



\end{thebibliography}
\end{document}